\documentclass[prd,twocolumn]{revtex4}
\begin{document}

\title{Slow-roll versus stochastic slow-roll inflation}

\author{Z. Haba\\Institute of Theoretical Physics, University of Wroclaw,
50-204 Wroclaw} \email{zbigniew.haba@uwr.edu.pl}\date{\today
}
\begin{abstract}We consider
   the classical wave equation with a thermal and  Starobinsky-Vilenkin
 noise which in the slow-roll and long wave approximation
describes the quantum fluctuations of the gravity-inflaton system
in an expanding metric. We investigate the resulting consistent
stochastic Einstein-Klein-Gordon system in the slow-roll regime.
We show in some models that the slow-roll requirements (of the
negligence of $\partial_{t}^{2}\phi$ )  can be satisfied in the
probabilistic sense for the stochastic system with quantum and
thermal noise for arbitrarily large time and an infinite range of
fields. We calculate  expectation values of some inflationary
variables taking into account quantum and thermal noise. We show
that the mean acceleration $\langle\partial_{t}^{2}a\rangle $ can
be negative or positive (depending on the model) when the random
fields take values beyond the classical range of inflation.
\end{abstract}

\maketitle
\section{Introduction}

In the standard classical approach to inflation the slow-roll
approximation is very useful. Fortunately, it applies just in the
range of field configuration where it is  needed,i.e., during the
time of the accelerated expansion. However, in the description of
the early stages of the universe evolution quantum and thermal
fluctuations are unavoidable. With the fluctuations the usual
criteria of the slow-roll strictly speaking do not apply. In fact
such fluctuations can lead to an eternal inflation, i.e. the
inflation never ends as suggested first by Linde
\cite{lindeinflation}. We are going to investigate the quantum and
thermal fluctuations quantitatively on the basis of differential
equations which are supposed to describe the exact time evolution
at least at the early stages. Our approach is based on the
Starobinsky approach \cite{starobinsky} to quantum fluctuations
and the description of thermal fluctuations known from the theory
of Brownian motion \cite{nelson} and applied to inflation in
\cite{bererafang}\cite{warm}\cite{power1}\cite{power2}\cite{habaejc}.
Starobinsky \cite{starobinsky} suggested an approach which treats
quantum scalar field non-perturbatively in classical Einstein
equations. The idea is based on an earlier observation
\cite{star1}\cite{star2}\cite{star3}\cite{vilenkin}that
the quantum scalar field in an expanding universe behaves as a
classical diffusion process. In such a case we obtain a stochastic
Einstein-Klein-Gordon (EKG)system. In the slow-roll limit the
stochastic EKG system is approximated by a first order
differential equation of a diffusion process. The approximation
applies well for deterministic systems in the range of slow-roll.
If we restrict the range of evolution of $\phi$ in the stochastic
equation in order to cut it to the classical range of inflation
then we must introduce boundary conditions as is done in
\cite{venn}\cite{venn3}. However, the stochastic equation
 makes sense in an unbounded  field configuration space till a random explosion time.
If we calculate this conditional probability distribution then it
satisfies the usual Fokker-Planck equation. We prefer the approach
based on an unbounded field evolution. On the basis of the
Fokker-Planck equation we study the problem of probabilistic
estimates of the error of the slow-roll approximation. We hope
that as in the theory of the Brownian motion of a particle in the
medium with a friction the first order equation well approximates
the second order stochastic non-linear oscillator equation
\cite{nelson}.

The plan of the paper is the following. In sec.2 we recall the
basic equations of the inflaton model. We describe the standard
slow-roll approximation in sec.3. The diffusion process
corresponding to slow-roll approximation is discussed in sec.4. In
sec.5 we describe  some models of the inflaton potential and
calculate the scale factor as a function of the field. In sec.6
the slow-roll approximation is discussed on the basis of the
probability distribution of the inflaton and its time derivatives.
The aim is to set a framework which allows to estimate corrections
to the slow-roll. In sec.7 we review standard results of the
theory of diffusion processes in order to explain what happens
with solutions of the inflaton models. In sec.8 we  calculate the
stationary probability distribution of the inflaton diffusion
process. In the main sec.9 of this paper we calculate some large
time  expectation values using the stationary probability
distribution  (this section can be considered as an application of
the methods and results of \cite{habad}).
 In the Appendix we explain the minor changes which
come out from a change of time (from cosmic time to the e-fold
time).

\section{Inflaton wave equation}
In the standard approach to inflation \cite{inflation} we consider
a flat expanding background metric
\begin{displaymath}
ds^{2}=g_{\mu\nu}dx^{\mu}dx^{\nu}=dt^{2}-a^{2}d{\bf x}^{2}
\end{displaymath}
Accelerated expansion is generated by the scalar field satisfying
a non-linear wave equation (where $H=a^{-1}\partial_{t}a$)

\begin{equation}
\partial_{t}^{2}\phi-a^{-2}\triangle\phi+3H\partial_{t}\phi+V^{\prime}(\phi)=0
 \end{equation}
 The expansion scale $a$ is determined by the Friedman equation

 \begin{equation}
 H^{2}=\frac{8\pi G}{3}(V+\frac{1}{2}(\partial_{t}\phi)^{2})
\end{equation}
We are going  to neglect $\partial_{t}^{2}\phi$ in eq.(1). Hence,
we require $\vert\partial_{t}^{2}\phi\vert<<\vert
3H\partial_{t}\phi\vert$. This condition is satisfied if
\begin{equation}\tilde{\epsilon}=\frac{1}{16\pi G}
(V^{\prime})^{2}V^{-2}\end{equation}and \begin{equation}
\tilde{\eta}=\frac{1}{8\pi G}V^{\prime\prime}V^{-1}
\end{equation}
are small. We calculate\begin{equation}
\partial_{t}^{2}a=\frac{a^{2}H^{3}}{\partial_{t}a}(1-\tilde{\epsilon})=aH^{2}(1-\tilde{\epsilon})
\end{equation}(differentiating eq.(2) with the
neglect of $\partial_{t}^{2}\phi$). We have an acceleration if
$\tilde{\epsilon}<1$.
\section{Stochastic
Einstein-Klein-Gordon system}

 We consider a modified version of eq.(1)
\begin{equation}
\partial_{t}^{2}\phi-a^{-2}\triangle\phi+
(3H+\beta\gamma^{2})\partial_{t}\phi+V^{\prime}(\phi)+\frac{3}{2}\beta\gamma^{2}H\phi=\eta.
 \end{equation}
Eq.(6) is to describe the inflaton in the long wave limit and
includes a noise $\eta$ resulting from quantum fluctuations of
refs.\cite{starobinsky}\cite{star1}-\cite{vilenkin} as well as
from the random interaction with the environment at the temperature
$\beta^{-1} $ introduced in \cite{berrera}\cite{adv} ( $\beta$ has
been omitted in \cite{adv}). $\beta\gamma^{2}$ is the friction
coming from the environment ( in \cite{berrera} $\beta\gamma^{2}$
is denoted as $\gamma^{2}$). In \cite{berrera} eq.(6) is derived
under different assumptions (especially in Appendix B of
ref.\cite{berrera}) than in \cite{adv}. With our assumptions
concerning the masses and couplings of the environmental fields
\cite{adv} we obtain eq.(6) in the long wave limit (the short wave
limit of the environmental noise in an expanding universe will be
discussed in \cite{habaprep}). With both random fluctuations the
noise $\eta$ on the rhs of eq.(6) takes the form
\begin{equation}
\eta\equiv \partial_{t}\xi=\gamma
a^{-\frac{3}{2}}\partial_{t}B+\frac{3}{2\pi}H^{\frac{5}{2}}\partial_{t}W
\end{equation}
where the first term is the thermal noise. The second term
describes the diffusive behaviour of the high frequency part of
the inflaton in an expanding universe. The factor
$a^{-\frac{3}{2}}$ in the thermal noise comes from the metric
weight $\det\vert g_{\mu\nu}\vert ^{\frac{1}{4}}$. The factor
$\frac{3}{2\pi}H^{\frac{5}{2}}$ is chosen in order to reproduce
the correlation functions of the quantum scalar field in an
expanding universe \cite{starobquan}. We express the thermal noise
by the Brownian motion $B$. This is a continuous Gaussian process
whose derivative has  the covariance
\begin{equation}
\langle
\partial_{t}B\partial_{s}B\rangle=\delta(t-s)\end{equation}
$\partial_{t}W$ (in eq.(7)) is an independent Gaussian stochastic
process with the same covariance (8).

With the thermal noise we would violate the conservation law for
the scalar energy-momentum $T_{\mu\nu}(\phi)$. If
$(T^{0\nu}(\phi))_{;\nu}=Q$ then we introduce a compensating dark energy
density $\rho_{d}$ such that $\partial_{t}\rho_{d}+3H(w+1)\rho_{d}=-Q$  with
  $w=-1$ (then $\rho_{d}=-\int dt Q$ see \cite{habad} for more details). Then, eq.(2) still remains true in a differential form.  In such a case we obtain a closed
random dynamical system (in the first order form) with the
Starobinsky-Vilenkin noise $W$ and the environmental noise  $B$
\cite{habad}
\begin{equation} d\phi=\Pi dt ,
\end{equation}
\begin{equation}\begin{array}{l}
d\Pi=-(3H+\beta\gamma^{2})\Pi dt -V^{\prime}dt
-\frac{3}{2}\beta\gamma^{2}H\phi dt \cr+\gamma a^{-\frac{3}{2}}\circ
dB+\frac{3}{2\pi}H^{\frac{5}{2}}\circ dW  ,
\end{array}\end{equation}
\begin{equation}
dH=-4\pi G\Pi^{2}dt   ,
\end{equation}\begin{equation}
da=H adt .
\end{equation} with eq.(11) replacing eq.(2).
We interprete the stochastic equations in the Stratonovitch sense
and use the $\circ$-notation for Stratonovitch differentials
following ref.\cite{ikeda}. From eq.(12)
\begin{equation}
\ln(\frac{a}{a_{0}}) =\int_{0}^{t}Hds.
\end{equation}

\section{Diffusion  equation for the slow-roll
inflation} The program outlined at the end of  sec.3 to solve the
stochastic Einstein-Klein-Gordon (EKG) equations for the
$(a,H,\phi,\Pi)$ variables and calculate the probability
distribution is difficult to perform in a non-perturbative way. We
shall rely on approximations. On the preliminary stage we neglect
the spatial derivatives in eq.(1). We omit the noise in eq.(10)
and the $\beta\gamma^{2}$ terms. Then, it follows from eqs.(9)-(12)
that
\begin{equation}
H=\sqrt{\frac{8\pi G}{3}(V+\frac{1}{2}\Pi^{2})}.\end{equation} and
\begin{equation}a=a_{0}\exp\Big(-8\pi G\int d\phi V(V^{\prime})^{-1}\Big)
\end{equation}
Using, eqs.(9)-(12) we can obtain the Fokker-Planck equation for
the probability distribution of $(a,\phi,\Pi)$. On a formal level
we make the next simplifying assumption

 $d\Pi\simeq 0$ and
$\gamma\simeq 0$ in eqs.(9)-(12). We  justify this approximation
(if $H$ is large) in sec.5 on the level of probability
distributions  . Then,  the system of eqs.(9)-(12) is reduced to

\begin{equation}
d\phi = -\frac{V^{\prime}}{3H}dt +\frac{\gamma}{3H}
a^{-\frac{3}{2}}\circ dB+\frac{1}{2\pi}H^{\frac{3}{2}}\circ dW,
\end{equation}
with explicit functions $H(\phi)$ and $a(\phi)$. The approximation
of the stochastic non-linear  wave equation (6) by a diffusion
equation has been extensively discussed in the theory of Brownian
motion \cite{nelson}. The main arguments for this approximation
will be discussed in sec.5.

 Eq.(16) in the Stratonovich
interpretation of the stochastic equations \cite{ikeda} leads to
the equation (retrospective Kolmogorov equation) for the
transition function $P(t,\phi;s,\phi^{\prime})$ \cite{gikhman}
\begin{equation}\begin{array}{l}
\partial_{t}P=\frac{\gamma^{2}}{18}\frac{1}{Ha^{\frac{3}{2}}}\partial_{\phi}\frac{1}{Ha^{\frac{3}{2}}}\partial_{\phi}P
+\frac{1}{8\pi^{2}}H^{\frac{3}{2}}\partial_{\phi}H^{\frac{3}{2}}\partial_{\phi}P
\cr-(3H)^{-1}V^{\prime}\partial_{\phi}P
.\end{array}\end{equation}and to the adjoint (prospective
Kolmogorov or Fokker-Planck) equation for the probability
distribution $P(s,\phi^{\prime};t,\phi)$
\begin{equation}\begin{array}{l}
\partial_{t}P=\frac{\gamma^{2}}{18}\partial_{\phi}\frac{1}{Ha^{\frac{3}{2}}}\partial_{\phi}\frac{1}{Ha^{\frac{3}{2}}}P
+\frac{1}{8\pi^{2}}\partial_{\phi}H^{\frac{3}{2}}\partial_{\phi}H^{\frac{3}{2}}P
\cr+\partial_{\phi}(3H)^{-1}V^{\prime}P .\end{array}\end{equation}

\section{The evolution of the scale factor  $a(\phi)$ }  The dependence of $a$ on $\phi$ is  determined  in the
slow-roll approximation by eq.(15).  Let us consider some examples
of potentials appearing in inflation models
\cite{ency}\cite{starpot0}\cite{starpot}\cite{natural}.
 For a chaotic inflation \cite{lindeinflation}  $V=g\phi^{2n}$ ($n\geq 1)$

\begin{equation}
a=a_{0}\exp\Big(-2\pi Gn^{-1}\phi^{2}\Big).
\end{equation}
If  $V=g\exp(\lambda\phi) $ then
\begin{equation}
a=a_{0}\exp\Big(-\frac{8\pi G}{\lambda}\phi\Big).\end{equation}
 If $\phi\rightarrow +\infty$ then
$a\rightarrow 0$, if $\phi\rightarrow -\infty$ then $a\rightarrow
\infty$. For the Starobinsky potential \cite{starpot}
\begin{equation} V=g(\exp(\lambda\phi)-1)^{2}
\end{equation}
we have
\begin{equation}
a=a_{0}\exp\Big(-\frac{4\pi G}{\lambda}\phi-\frac{4\pi
G}{\lambda^{2}}\exp(-\lambda \phi)\Big)
\end{equation}
$a(\phi)$ is bounded and $a\rightarrow 0$ if $\vert
\phi\vert\rightarrow \infty$.

The potential for the  " natural inflation" \cite{natural}
describing the axion inflation can be defined as
\begin{equation}
V=g (1+\cos(\lambda\phi))
\end{equation}
 From eq.(15) we obtain
\begin{displaymath}
a=a_{0}\vert \sin(\frac{\lambda\phi}{2})\vert^{\frac{16\pi
G}{\lambda^{2}}}
\end{displaymath}
Here, $-\pi\leq \lambda\phi\leq\pi$.

 For the double-well potential
\begin{equation}
V(\phi)=\frac{g}{4}(\phi^{2}-\frac{\mu^{2}}{g})^{2} .
\end{equation}
\begin{equation}\begin{array}{l}
a=a_{0}\vert\phi\vert^{\frac{2\pi G\mu^{2}}{g}}\exp(-\pi
G\phi^{2}) .
\end{array}\end{equation}

When $\vert\phi\vert\rightarrow\infty$ then $a\rightarrow 0$.

 In subsequent
sections we also consider the double-well in the form
\begin{equation} V=\frac{g_{n}}{4}(\phi^{2}-\frac{\mu^{2}}{g})^{n}
\end{equation}(where $g_{n}$ is a dimensional coupling constant)
then
\begin{equation}
a=a_{0}\vert\phi\vert^{\frac{4\pi G\mu^{2}}{ng}}\exp(-\frac{2\pi
G}{n}\phi^{2})
\end{equation}
\section{Slow-roll approximation for the probability distribution}
The approximation of eq.(6) by eq.(16) can be expressed as an
omission of $\partial_{t}^{2}\phi$. For a deterministic system
this approximation can be controlled by a solution of the
slow-roll equation. In a stochastic system solely estimates of
probabilities make sense. We must estimate the probability that
$\partial_{t}^{2}\phi$ is small.In the system of differential
equations (9)-(12) $a$ is a random variable. In principle, we
could write down a partial differential equation for the
probability distribution $P$ of $(a,\phi,\Pi)$ and subsequently
integrating it over $a$ and $\Pi$ we could reduce it to the
probability distribution of $\phi$ (this is the probability of
finding $\phi$ no matter what are the values of $a$ and $\Pi$).
This is however technically difficult to achieve. We follow the
approximations of secs.4-5. We assume that
 $H$ (14) and $a$ (15) are explicit functions of $\phi$ and $\Pi$.
  The
transition function
$P^{ret}_{t}(\phi,\Pi;\phi^{\prime},\Pi^{\prime})$ of the process
(9)-(10) (neglecting the $\beta\gamma^{2}$ terms, i.e., assuming
$\beta\gamma^{2}<<3H$) satisfies the equation (retrospective Kolmogorov
equation)\begin{equation}\begin{array}{l}
\partial_{t}P^{ret}=\frac{\gamma^{2}}{2}a^{-\frac{3}{2}}\partial_{\Pi}a^{-\frac{3}{2}}\partial_{\Pi}P^{ret}
\cr+\frac{9}{8\pi^{2}}H^{\frac{5}{2}}\partial_{\Pi}H^{\frac{5}{2}}\partial_{\Pi}P^{ret}
-(3H\Pi+V^{\prime})\partial_{\Pi}P^{ret}
+\Pi\partial_{\phi}P^{ret}.\end{array}\end{equation} The adjoint
(Fokker-Planck) equation for
$P_{t}(\phi^{\prime},\Pi^{\prime};\phi,\Pi)$ is
\begin{equation}\begin{array}{l}
\partial_{t}P=\frac{\gamma^{2}}{2}\partial_{\Pi}a^{-\frac{3}{2}}\partial_{\Pi}a^{-\frac{3}{2}}P
\cr+\frac{9}{8\pi^{2}}\partial_{\Pi}H^{\frac{5}{2}}\partial_{\Pi}H^{\frac{5}{2}}P
+\partial_{\Pi}(3H\Pi+V^{\prime})P -\partial_{\phi}\Pi
P.\end{array}\end{equation} In eq.(28)we make a change of
variables $(\phi,\Pi)\rightarrow (X,Y)$ where
\begin{equation}
X=\phi+\frac{1}{3H}\Pi
\end{equation}
\begin{equation}
Y=\Pi
\end{equation}
In new coordinates the transition function (28) reads
\begin{equation}\begin{array}{l}
\partial_{t}P^{ret}=\frac{\gamma^{2}}{2}a^{-\frac{3}{2}}(\partial_{Y}+\frac{1}{3H}\partial_{X})a^{-\frac{3}{2}}(\partial_{Y}+\frac{1}{3H}\partial_{X})P^{ret}
\cr
+\frac{9}{8\pi^{2}}H^{\frac{5}{2}}(\partial_{Y}+\frac{1}{3H}\partial_{X})H^{\frac{5}{2}}(\partial_{Y}+\frac{1}{3H}\partial_{X})
P^{ret} \cr
-(3HY+V^{\prime})(\partial_{Y}+\frac{1}{3H}\partial_{X})P^{ret}
+Y(1+Y\partial_{\phi}(\frac{1}{3H}))\partial_{X}P^{ret}.\end{array}\end{equation}
The change of coordinates in the Fokker-Planck equation (29) is
\begin{equation}\begin{array}{l}
\partial_{t}P=\frac{\gamma^{2}}{2}(\partial_{Y}+\frac{1}{3H}\partial_{X})a^{-\frac{3}{2}}(\partial_{Y}+\frac{1}{3H}\partial_{X})a^{-\frac{3}{2}}P
\cr
+\frac{9}{8\pi^{2}}(\partial_{Y}+\frac{1}{3H}\partial_{X})H^{\frac{5}{2}}(\partial_{Y}+\frac{1}{3H}\partial_{X})H^{\frac{5}{2}}
P \cr +(\partial_{Y}+\frac{1}{3H}\partial_{X})(3HY+V^{\prime})P
-\partial_{X}\Big(Y(1+Y\partial_{\phi}(\frac{1}{3H}))P\Big).\end{array}\end{equation}
We wish to get rid off the dependence of $P$ on $Y$ altogether. We
impose the slow-roll requirement on the probability distribution
$P$ assuming that the dependence (14) of $H$ on $\Pi$ is
negligible and we neglect $\partial_{\phi}\frac{1}{H} $ in
eqs.(32)-(33)  (the assumption that $H$ is slowly varying). In
eqs.(28)-(29) the order of derivatives over $\Pi$ and
multiplicative factors  is not relevant when the dependence of $H$
and $a$ on $\Pi$ is neglected. However, after a change of
variables  the multiplicative factors become dependent on $X$ and
$Y$. When we subsequently perform a limit of large $H$ (slow-roll)
then the order of factors becomes significant. So, the
approximation has to be performed in a consistent way. In these
equations we approximate
$V^{\prime}(\phi)=V^{\prime}(X-\frac{1}{3H}Y)\simeq
V^{\prime}(X)$, $H(\phi)=H(X-\frac{1}{3H}Y)\simeq H(X)$ and
neglect $\partial_{\phi}\frac{1}{3H}$. We integrate the transition
function over $Y$,i.e., $\int dY P^{ret}(s,X,Y;t,
X^{\prime},Y^{\prime})\equiv \hat{P}^{ret}(s,X;t,
X^{\prime},Y^{\prime})$. Then, the limit of slowly varying $H$ and
small $\frac{Y}{H}$ for the integrated (over $Y$ ) transition
function is
\begin{equation}\begin{array}{l}
\partial_{t}\hat{P}^{ret}=\frac{\gamma^{2}}{2}a^{-\frac{3}{2}}\frac{1}{3H}\partial_{X}a^{-\frac{3}{2}}\frac{1}{3H}\partial_{X}\hat{P}^{ret}
\cr+\frac{9}{8\pi^{2}}H^{\frac{5}{2}}\frac{1}{3H}\partial_{X}H^{\frac{5}{2}}\frac{1}{3H}\partial_{X}
\hat{P}^{ret}
-V^{\prime}\frac{1}{3H}\partial_{X}\hat{P}^{ret}-3H\hat{P}^{ret}.\end{array}\end{equation}

 The probability
distribution of $X$ is

 $\hat{P}_{t}=\int dY
P(s,X^{\prime},Y^{\prime};t, X,Y)$. It follows that the integral
over $Y$ of the Fokker-Planck equation (33) gives

\begin{equation}\begin{array}{l}
\partial_{t}\hat{P}_{t}=\frac{\gamma^{2}}{2}\frac{1}{3H}\partial_{X}a^{-\frac{3}{2}}\frac{1}{3H}\partial_{X} a^{-\frac{3}{2}}\hat{P}_{t}
\cr+\frac{9}{8\pi^{2}}\frac{1}{3H}\partial_{X}H^{\frac{5}{2}}\frac{1}{3H}\partial_{X}H^{\frac{5}{2}}
\hat{P}_{t}+\frac{1}{3H}\partial_{X}(V^{\prime}\hat{P}_{t}).\end{array}\end{equation}
This equation is different from  the Fokker-Planck equation (18)
which we get from the stochastic equation (16). The difference
concerns the terms $\partial_{\phi}\frac{1}{3H}$ which must be
neglected in a consistent way. An alternative procedure is to
define $\tilde{P}=a^{-3}P^{ret}$ insert it in eq.(34) (for the
integrated transition function) and take the adjoint of this
equation. As a result we obtain

\begin{equation}\begin{array}{l}
\partial_{t}\tilde{P}=\frac{\gamma^{2}}{2}\partial_{X}a^{-\frac{3}{2}}\frac{1}{3H}\partial_{X}
a^{-\frac{3}{2}}\frac{1}{3H}\tilde{P}
+\frac{9}{8\pi^{2}}\partial_{X}H^{\frac{5}{2}}\frac{1}{3H}\partial_{X}H^{\frac{5}{2}}\frac{1}{3H}
\tilde{P} \cr
+\partial_{X}(V^{\prime}\frac{1}{3H}\tilde{P})\end{array}\end{equation}
which coincides with eq.(18).

In this way we have achieved the slow-roll approximation of the
stochastic system (9)-(12) on the basis of probability
distributions. The approximation could be controlled by estimating
the neglected terms of the exact probability distributions
(28)-(29).
\section{Application of some general results on diffusions}
We consider a diffusion equation of the form
\begin{equation}
d\xi=-\beta(\xi)dt+\sigma(\xi)\circ dB=-\tilde{\beta}dt +\sigma dB
\end{equation}
where $\sigma dB $ is Ito differential \cite{ikeda} and
\begin{displaymath}
\tilde{\beta}=\beta-\frac{1}{2}\sigma^{\prime}\sigma
\end{displaymath}
In general, solutions of eq.(37) exist till the random explosion
time $\tau$. Nevertheless, probability distributions of sec.6
(satisfying the Kolmogorov equations) are well defined for
arbitrary time as conditional probabilities $P_{t}(A)\equiv
P(\xi_{t}\in A\vert t<\tau(\xi))$. There is a useful criterion
\cite{khasminski} when the stochastic process $\xi$ can be defined
for arbitrarily large time. Define a Lyapunov function $L(x)$ as a
twice differentiable positive function increasing to $+\infty$
when $\vert x\vert\rightarrow \infty$ and satisfying the
inequality
\begin{equation}
{\cal A}L< -\alpha L+r
\end{equation}
where $\alpha$ and $r$ are positive constants and ${\cal A}$ is
the differential operator on the rhs of the retrospective
Kolmogorov equation. The basic theorem of diffusion processes
states \cite{khasminski} that if for the given diffusion process
(defined by ${\cal A}$) there exists a Lyapunov function then the
process can be defined for arbitrarily large time.

Let us consider as a simple example $\sigma=const$. Take
$L=x^{2}$, then for the process (37)
\begin{displaymath}
{\cal A}L=\sigma^{2}-2\beta x
\end{displaymath}
If $2\beta x \geq \alpha x^{2}-r$ with $r\geq 0$ then the
inequality (38) is satisfied. It holds true if $\beta= V^{\prime}$
with $V=x^{2n}$ and fails for $V=x^{2n-1}$ (natural $n$).

Next, we can calculate the asymptotic probability distribution. If
$\tilde{\beta }$ is a continuous function, $\sigma$ is
continuously differentiable and $\sigma^{2}>0$ then the
Fokker-Planck equation has a unique solution. If $\int f_{*}=1$
where
\begin{equation}
f_{*}=K \sigma^{-2}\exp(-\int 2\tilde{\beta}\sigma^{-2})
\end{equation} with a certain finite $K>0$
then for any bounded normalized $f\geq 0$ there exists the unique
limit $f_{*}$ when $t\rightarrow \infty$ of the Markov semigroup
$P_{t}f$ \cite{khasminski}\cite{mackey}. This $f_{*}$ is called
the stationary probability distribution . In our case
\begin{equation}\frac{1}{2}\sigma^{2}=\frac{1}{18}\frac{\gamma^{2}}{a^{3}H^{2}}
+\frac{1}{8\pi^{2}}H^{3}
\end{equation} and $\beta=\frac{1}{3H}V^{\prime}$.
In most models of inflation if $\gamma^{2}>0$ (the  thermal noise
 is non-zero) then  $\sigma ^{2}>0$. If there is no thermal noise then $\sigma\simeq 0$ if
 $H^{2}\simeq V\simeq 0$ what happens  in the examples of
 sec.5.  Hence, without the thermal noise
 there may be difficulties with the limit  $t\rightarrow \infty$.

\section{Stationary probability distribution of the inflaton}
 The stationary probability $P_{\infty}(\phi)$ is the limit of $P_{t}$ for $t\rightarrow\infty$
 \cite{mackey}.
  In
 the Stratonovitch interpretation it can be obtained from the
 requirement $\partial_{t}P_{\infty}=0$ which gives (after an integration
 over $\phi$)\begin{equation}\begin{array}{l}
\frac{\gamma^{2}}{18}\frac{1}{Ha^{\frac{3}{2}}}\partial_{\phi}\frac{1}{Ha^{\frac{3}{2}}}P_{\infty}
+\frac{1}{8\pi^{2}}H^{\frac{3}{2}}\partial_{\phi}H^{\frac{3}{2}}P_{\infty}
+(3H)^{-1}V^{\prime}P_{\infty} =0.\end{array}\end{equation}
The stationary solution of eq.(41) without the
Starobinsky-Vilenkin noise is (from eq.(39))
\begin{equation}\begin{array}{l}
 P_{\infty}=\sqrt{V}\exp\Big(-12\pi
G\int^{\phi}d\phi^{\prime}
(V^{\prime})^{-1}V\Big)\cr\exp\Big(-\frac{6}{\gamma^{2}}\sqrt{\frac{8\pi
G}{3}}\int d\phi V^{\prime}\sqrt{V}\exp(-24\pi
G\int^{\phi}d\phi^{\prime} (V^{\prime})^{-1}V)\Big),
\end{array}\end{equation} where the exponential factors in eq.(42) come
from the formula for $a^{3}$ (eq.(15)).
 For a large $\vert\phi\vert$  we have $a(\phi)\rightarrow 0$ in most of our models of
 sec.5. Hence, we can neglect the $\gamma^{-2}$ term. Then,
\begin{equation}\begin{array}{l}
 P_{\infty}\simeq \sqrt{V}\exp\Big(-12\pi
G\int^{\phi}d\phi^{\prime} (V^{\prime})^{-1}V\Big)\end{array}
\end{equation}
If $\gamma=0$ (the thermal noise is absent) then we obtain the
Starobinsky solution (discussed also by Vilenkin \cite{vilenkin}
and Linde \cite{linde}) \begin{equation}
P_{\infty}=V^{-\frac{3}{4}}\exp(\frac{3}{8G^{2}}\frac{1}{V}).
\end{equation} The solution (44) is not normalizable if $V\geq 0 $ and $V=0$ at a certain
$\phi$ or $V$ does not decay fast enough for large $\phi$. The
authors \cite{venn}\cite{venn3} impose boundary conditions
excluding the regions where $P_{\infty}$ (44) is not integrable.
Then, one has to study whether the dependence on boundary
conditions has some consequences on calculated expectation values.

With $\gamma\neq 0$ in eq.(41) we write
\begin{equation}
\tilde{P}=H^{-1}a^{-\frac{3}{2}}P_{\infty} .
\end{equation}
Then, the equation for $\tilde{P}$ is
\begin{displaymath}
\frac{\gamma^{2}}{18}H^{-1}a^{-\frac{3}{2}}\partial_{\phi}\tilde{P}+\frac{1}{8\pi^{2}}H^{\frac{3}{2}}\partial_{\phi}(H^{\frac{5}{2}}a^{\frac{3}{2}}\tilde{P})=
-\frac{1}{3}V^{\prime}a^{\frac{3}{2}}\tilde{P}.\end{displaymath}
Using the formulas for $H$ and for  $a$ (eq.(15)) we obtain
\begin{equation}\begin{array}{l}
\ln\tilde{P}=-6\int d\phi H a^{3}
(\gamma^{2}+\frac{9}{4\pi^{2}}H^{5}a^{3})^{-1}\cr\Big(V^{\prime}+\frac{10}{3}G^{2}VV^{\prime}
-32\pi G^{3}V^{3} (V^{\prime})^{-1}\Big).\end{array}
\end{equation}If $V^{\prime}\geq 0$ and $\Big(1+\frac{10}{3}G^{2}V
-32\pi G^{3}V^{3} (V^{\prime})^{-2}\Big)\geq 0$ then $\tilde{P}$
improves integrability and we can use the saddle point method for
calculations. If $\ln \tilde{P}\geq \gamma^{-2}K $ with a certain
constant $K$ then for any $f$ we have

 $\vert\int f P_{\infty}\vert
\leq \exp(\gamma^{-2}K)\int \vert f\vert Ha^{-\frac{3}{2}}$. This
estimate can be sufficient for integrability. In general, we have
to find a $\phi$-dependent upper bound on $\ln\tilde{P}$  in order
to prove integrability.

 For most potentials of sec.5 $a\rightarrow 0$ for
$\vert\phi\vert\rightarrow \infty$ then $a^{3}H^{5}\rightarrow 0$.
Hence, we get from eqs.(45)-(46) $P_{\infty}\simeq
Ha^{\frac{3}{2}}$ at large $\gamma$ coinciding with the formula
(43).  As an explicit example we consider the chaotic inflation
potential $V=g\phi^{2n}$ . The integrand $\ln\tilde{P} $ in eq.(46)
is a bounded function of $\phi$ because $a\simeq\exp(-2\pi
Gn^{-1}\phi^{2}) $.Then, the formula (45) gives for a large
$\vert\phi\vert$ (small $a$)
\begin{equation}
P_{\infty}\simeq\vert\phi\vert^{n}\exp(-3\pi
Gn^{-1}\phi^{2}) .
\end{equation} At small $\phi$  we have $a^{3} H^{5}\rightarrow 0$ . So, the formula
(43) is   applicable  for small as well as large $\phi$.
 The Starobinsky formula (44)
(obtained without the thermal noise) gives $P_{\infty}$ which is
not integrable at $\phi=0$.

When $a^{3}H^{5}\rightarrow \infty$ and we neglect $\gamma^{2}$ in
eq.(46) then
\begin{equation}
\ln\tilde{P}\simeq -\frac{8\pi^{2}}{3}\int d\phi H^{-4}
\Big(V^{\prime}+\frac{10}{3}G^{2}VV^{\prime} -32\pi G^{3}V^{3}
(V^{\prime})^{-1}\Big).
\end{equation}
We can calculate the rhs of eq.(48) and convince ourselves that in
this case we reach the Starobinsky formula (44). In the estimates
of integrals with respect to $P_{\infty}$ in the next section we
cannot let $\gamma^{2}\rightarrow 0$.

\section{Probabilistic estimates of the slow-roll regime}
We  would like to estimate on the basis of the Fokker-Planck
equation (18) the probability that $\vert\partial_{t}^{2}\phi\vert
<<\vert 3H\partial_{t}\phi\vert$. As a consequence of the
(deterministic) equations of sec.2 we can calculate
\begin{equation}
\langle\partial_{t}^{2}\phi(3H\partial_{t}\phi)^{-1}\rangle=\frac{1}{3}
\langle\tilde{\epsilon}-\tilde{\eta}\rangle.
\end{equation}
In the classical EKG equations of sec.3 the regime of small
$\tilde{\epsilon}$ and $\tilde{\eta}$ is closely related with the
inflation regime according to eq.(5). We can study both problems
in probabilistic terms by a calculation of  the mean values of the
acceleration and the derivatives of the inflaton (as in eq.(49)).
So, according to eq.(5) (strictly speaking this criterion applies
if the rhs of eq.(49) is  small)\begin{equation}
\langle\partial_{t}^{2}a\rangle=\langle aH^{2}\rangle -\langle
aH^{2}\tilde{\epsilon}\rangle.
\end{equation} The inequality
$\langle aH^{2}\rangle >\langle aH^{2}\tilde{\epsilon}\rangle$
depends on the probability distribution of $\phi$ determined by the stochastic equation (16).
However, without the deterministic inflation the stochastic approximation to the
quantum noise is hard to justify. One should rather consider eq.(50)
as a condition for the preservation of inflation by quantum and thermal fluctuations.

In order to calculate the probability distribution of the
variables  in eqs.(49)-(50) we would need to solve the stochastic
equations or the Fokker-Planck equations. We are able to calculate
only the mean values (49)-(50) at large time using the stationary
probability distribution. Nevertheless, such calculations give
some hints on the role of thermal and quantum fluctuations in  the
evolution of inflation. So, in the model $V=g \phi^{2n}$ ($n>1$)
with the thermal noise we have on the basis of the
approximation(47)
\begin{equation}\begin{array}{l}
\langle \tilde{\epsilon}\rangle=4n^{2}(16\pi G)^{-1} \int d\phi
\vert\phi\vert^{n-2}\exp (- \frac{3\pi G}{n}\phi^{2})\cr\Big(\int
d\phi \vert\phi\vert^{n}\exp (- \frac{3\pi
G}{n}\phi^{2})\Big)^{-1} =\frac{2n}{n-1}
\end{array}\end{equation}and
\begin{displaymath}
\langle\tilde{\eta}\rangle= \frac{2(2n-1)}{n-1}
\end{displaymath}
So, $\langle \tilde{\epsilon}\rangle$ and
$\langle\tilde{\eta}\rangle$ are never small. The reason is that
in the $\phi^{2n}$ models $\langle \tilde{\epsilon}\rangle$ and
$\langle\tilde{\eta}\rangle$ behave as $n^{2}\phi^{-2}$. The
probability distribution of the stochastic process $\phi_{t}$
gives big weight to small values of $\phi$. For this reason the
expectation values of $\langle \tilde{\epsilon}\rangle$ and
$\langle\tilde{\eta}\rangle$  cannot be small. As a consequence
the approximation on which the stochastic equation (16) is based
does not seem to be applicable. Nevertheless, if we still insist
on applying eq.(50) then for  large $n$ (owing to the term
$H^{2}\simeq \phi^{2n}$)
\begin{equation} \langle\partial_{t}^{2}a\rangle\simeq\langle
a\vert\phi\vert^{n}\rangle -\langle a\vert\phi\vert^{n}\tilde{\epsilon}\rangle >0
\end{equation}
because  the expectation value $\langle \vert\phi\vert^{r}\rangle$ is an
increasing function of $r$ .

 Let us consider next the double-well
potential (26). The Starobinsky formula (44) does not give an
integrable stationary probability because $\frac{1}{V}$ is
singular at $g\phi^{2}=\mu^{2}$. If the quantum noise is absent
then
\begin{equation}\begin{array}{l}
P_{\infty}=Ha^{\frac{3}{2}}\cr\exp\Big(-\gamma^{-2}g\sqrt{24\pi
Gg}\cr\int
d\phi\vert\phi^{2}-\frac{\mu^{2}}{g}\vert(\phi^{2}-\frac{\mu^{2}}{g})\phi
\vert\phi\vert^{\frac{6\pi G\mu^{2}}{g}}\exp(-3\pi
G\phi^{2})\Big)\end{array}\end{equation} where
\begin{equation} Ha^{\frac{3}{2}}=\sqrt{\frac{2\pi Gg}{3}}\vert\phi^{2}-\frac{\mu^{2}}{g}\vert \vert\phi\vert^{\frac{3\pi
G\mu^{2}}{g}}\exp(-\frac{3}{2}\pi G\phi^{2})
\end{equation}
for a large $\gamma$ the last factor in eq.(53) is irrelevant.
It can be seen thaat $\langle\tilde{\epsilon}\rangle$
and $\langle\tilde{\eta}\rangle$ are infinite because of the
singularity of the integral at $\phi^{2}=\frac{\mu^{2}}{g}$.
Nevertheless, we
can calculate for large $\gamma$
\begin{equation}\begin{array}{l}
\langle\partial_{t}^{2}a\rangle=\frac{8\pi
G}{3}\int\Big((\phi^{2}-\frac{\mu^{2}}{g})^{2}-\frac{\phi^{2}}{8\pi
G}\Big) \vert\phi\vert^{\frac{5\pi
G\mu^{2}}{g}}\exp(-\frac{5}{2}\pi G\phi^{2}) \cr \Big(\int
\vert\phi^{2}-\frac{\mu^{2}}{g}\vert\vert\phi\vert^{\frac{3\pi
G\mu^{2}}{g}}\exp(-\frac{3}{2}\pi G\phi^{2})\Big)^{-1}\end{array}
\end{equation}
 $ \langle\partial^{2}a\rangle$ is positive (because $\int x^{\alpha}\exp(-x^{2})$ is an
increasing function of $\alpha$). If we calculate the integral
(53) at $\gamma\rightarrow 0$ by means of the saddle point method
then there are saddle points at $\phi=0$ and at $\phi=\pm
\frac{\mu}{\sqrt{g}}$. The saddle points give zero to the first
term in eq.(50) (and the corresponding term in the integral (53)) and
non-zero contribution to the second term. Hence, acceleration can
be negative for a small $\gamma$.
$\langle\tilde{\epsilon}\rangle$ and $\langle\tilde{\eta}\rangle$
are finite in a model with the double-well (26)($n\geq 4$). Then,
we have
\begin{equation}
\tilde{\epsilon}=\frac{n^{2}}{4\pi
    G}\phi^{2}(\phi^{2}-\frac{\mu^{2}}{g})^{-2}
\end{equation}
    \begin{displaymath}
    \tilde{\eta}=2n\frac{(2n-1)\phi^{2}-\frac{\mu^{2}}{g}}{4\pi
        G}(\phi^{2}-\frac{\mu^{2}}{g})^{-2}
    \end{displaymath}

and for a large $\gamma$
\begin{equation}
P_{\infty}\simeq
\vert\phi^{2}-\frac{\mu^{2}}{g}\vert^{\frac{n}{2}}a=\vert\phi^{2}-\frac{\mu^{2}}{g}\vert^{\frac{n}{2}}\vert\phi\vert^{\frac{6\pi
G\mu^{2}}{ng}}\exp(-\frac{3\pi G}{n}\phi^{2})
\end{equation}

If $n\geq 4$ then the singularity of $\tilde{\epsilon}$ (and the
same singularity of $\tilde{\eta}$) are cancelled by the factor in
$P_{\infty}$ what makes the expectation values of
$\tilde{\epsilon}$ and $\tilde{\eta}$ finite.
For a large $\gamma$ we may use the approximate formula (57) for calculation of expectation values. Then
\begin{equation}\begin{array}{l}
 \langle\tilde{\epsilon}\rangle=\int dye^{-y}\vert y-\frac{3\pi G\mu^{2}}{ng}\vert^{\frac{n}{2}-2}y^{\frac{1}{2} +\frac{3\pi G\mu^{2}}{ng}}
\cr
\Big(\int dye^{-y}\vert y-\frac{3\pi G\mu^{2}}{ng}\vert^{\frac{n}{2}}y^{-\frac{1}{2} +\frac{3\pi G\mu^{2}}{ng}}\Big)^{-1}
\end{array}\end{equation}
For a large $\frac{ G\mu^{2}}{ng}$
we obtain
\begin{equation}
\langle\tilde{\epsilon}\rangle\simeq \frac{n^{2}g}{4\pi G\mu^{2}}
\end{equation} and a similar result for $\langle\tilde{\eta}\rangle$.
Hence, both  $\langle \tilde{\epsilon}\rangle$ and
$\langle\tilde{\eta}\rangle$ are small if $\gamma$ and $\frac{ G\mu^{2}}{ng}$ are large.

$\langle \tilde{\epsilon}\rangle$ and
$\langle\tilde{\eta}\rangle$ can also be small for small $\gamma$. In fact, for $n>4$ these
expectation values can be arbitrarily small when
$\gamma\rightarrow 0$ as can be seen from  calculations by the
saddle point method. It can also be shown that for $n\geq 4$ the
mean value of the acceleration is positive for large $\gamma$
(this result could be treated as an argument for an eternal
inflation \cite{lindeinflation}\cite{vilenkininflation})and
negative for small $\gamma$ .

As a next example (when the Starobinsky formula (44) does not give
a stationary probability)
 let us consider $V=g\exp(\lambda\phi)$.
If only the thermal noise is present then (according to eqs.(42)
and (22))
\begin{equation}\begin{array}{l}
P_{\infty}=\exp((\frac{\lambda}{2}-\frac{12\pi
G}{\lambda})\phi)\cr\exp\Big(-2\gamma^{-2}\sqrt{96\pi G g
}(3\lambda-\frac{48\pi
G}{\lambda})^{-1}\exp\Big((\frac{3}{2}\lambda-\frac{24\pi
G}{\lambda})\phi\Big)\Big)\end{array}
\end{equation}
$\int P_{\infty}$ is finite if  $\lambda^{2}
>24\pi G$. With both quantum and thermal
noises we must ensure for integrability  that  the last term in
the brackets (46) is positive. This will be the case if
\begin{displaymath}
(\frac{10}{3}\lambda G^{2}-\frac{32\pi
G^{3}}{\lambda})\exp(2\lambda\phi)>0
\end{displaymath}
Hence,  $\lambda^{2}>9,6 \pi G $ (ensures the previous requirement
$\lambda^{2}
>24\pi G$).
Then, $\tilde{\epsilon}=\frac{1}{16\pi G}\lambda^{2}>\frac{3}{2}$
and $\tilde{\eta}>3$ so that in eq.(49)
$\vert\tilde{\eta}-\tilde{\epsilon}\vert
>\frac{3}{2}$. The mean  acceleration (50) is negative for  $\lambda^{2}>24 \pi
G$ (for these values of  $\lambda^{2}$ it is also negative in the
exact deterministic model (9)-(12) \cite{powerlaw}). However,
because  $\tilde{\epsilon}$ and
 $\tilde{\eta}$ with the quantum noise are not small the application
 of the diffusive approximation to the quantum noise and at the next step
 a reduction of eq.(6) to eq.(16) cannot be justified.

In the case of the Starobinsky potential (21)\cite{starpot}
according to eq.(22)
\begin{equation}
P_{\infty}=\vert\exp(\lambda\phi)-1\vert\exp\Big(
 -\frac{6\pi G}{\lambda}\phi-\frac{6\pi
 g}{\lambda^{2}}\exp(-\lambda\phi)\Big)\exp(\ln\tilde{P})
 \end{equation}
 We obtain that $\ln \tilde{P}$ is bounded for negative $\phi$.
 For positive $\phi$ we have
 $\ln \tilde{P}<\frac{K}{\lambda\gamma^{2}}\phi$ with $K>0$. Hence, $P_{\infty}$ is
 integrable for a sufficiently large $\lambda$ (depending on $\gamma$).
 $\langle\tilde{\epsilon}\rangle$ and $\langle \tilde{\eta}\rangle$
 are infinite. $\langle\partial_{t}^{2}a\rangle$ is finite and
 negative for large $\lambda$. We  get finite $ \langle\tilde{\epsilon}\rangle$ and $\langle \tilde{\eta}\rangle$
 if instead of the Starobinsky potential we  take
 \begin{equation}V=g(\exp(\lambda\phi)-1)^{n}\end{equation} with $n\geq 4$.

 For the
natural inflation (23) the Starobinsky stationary distribution
(44) does not exist because of the singularity of $\frac{1}{V}$ at
$\lambda\phi =\pi$. The stationary probability distribution
(45)-(46) is well-defined and integrable because $a^{3}H^{5}$ is
bounded. $\langle \tilde{\epsilon}\rangle$ and $\langle
\tilde{\eta}\rangle$ are infinite as a consequence of the
singularity of $\tilde{\epsilon}$ and $ \tilde{\eta}$ at
$\lambda\phi =\pi$. The acceleration of eq.(50) is (up to a
positive normalization constant)
\begin{equation}\begin{array}{l}
\langle \partial^{2}a\rangle=\frac{8\pi gG}{3}\cr
\int_{-\frac{\pi}{\lambda}}^{\frac{\pi}{\lambda}}d\phi
\Big(\vert\sin(\frac{\lambda\phi}{2})\vert^{\frac{24\pi
G}{\lambda}} \cr- (1+\frac{\lambda^{2}}{96\pi
G})\vert\sin(\frac{\lambda\phi}{2})\vert^{\frac{24\pi
G}{\lambda}+2}\Big)\exp(\ln \tilde {P} ) \end{array}\end{equation}
where we applied the notation
 of eq.(46). $\ln\tilde{P}$ is a regular bounded function
 hence its contribution to the integral is  negligible (at large $\gamma$).
The $P_{\infty} $ integral of powers of $\sin^{2\mu-1}$  is equal
to $2^{2\mu-2}\Gamma(\mu)^{2}(\Gamma(2\mu))^{-1}$. Hence, $\langle
\partial^{2}a\rangle$ is negative for large $\lambda$. It can be seen from our calculations that if  instead
of the potential (23) we considered
\begin{equation}
V=g(1+\cos(\lambda\phi))^{n}
\end{equation}
with $n\geq 2$ then expectation values of $\tilde{\epsilon}$ and
$\tilde{\eta}$ would   be finite as follows from the approximation
(43) for the stationary probability distribution
\begin{equation}
P_{\infty}\simeq \vert \cos(\frac{\lambda\phi}{2})\vert^{n}\vert
\sin(\frac{\lambda\phi}{2})\vert^{\frac{24\pi G}{\lambda^{2}}}
\end{equation}
Then the singularity of $\tilde{\epsilon}$ and $\tilde{\eta}$
cancels with the corresponding factor in $P_{\infty}$.
  \section{Summary and conclusions}
  The slow-roll approximation is a useful tool in the
  study of deterministic as well as stochastic  EKG systems.
  In the deterministic case we stop using it as soon as
  the parameters (functions of $\phi$) $\tilde{\epsilon}$ and $\tilde{\eta}$
  become large. In the stochastic version we should argue that
  we stop applying slow-roll for a time $t$ such that the
  probability that e.g.$\tilde{\epsilon}>\frac{1}{2}$ and $\tilde{\eta}>\frac{1}{2}$
  becomes large. This time is difficult to estimate. As in the
  theory of Brownian motion \cite{nelson} we suggest that with the
  strong "friction" $H$ the first order equation (16) may be a reliable
  approximation to the second order differential equation (6).
  However, for a stochastic approximation of the quantum noise the
  requirement of small  $\tilde{\epsilon}$ and $\tilde{\eta}$ seems unavoidable.
  We distinguish a class of models where the expectation values of these inflation parameters are indeed small.
   We calculate  expectation values  of $\tilde{\epsilon}$ , $\tilde{\eta}$
   and $\partial_{t}^{2}a$  with respect to the stationary probability which gives
   the probability distribution for a large time. We show that in some models the
   acceleration can be positive showing that quantum and thermal fluctuations do
   not destroy inflation even at large time. The sign of the
acceleration can depend on the friction as we have shown in the
double-well model.

\section{Appendix:e-fold time} In the calculation of the
power spectrum \cite{mar}\cite{starven} in order to take into
account fluctuations of the quantum gravitational field the use of
the e-fold time $\nu$ defined by
\begin{equation}
\nu=\int Hdt
\end{equation}
is crucial. The change of time does not have substantial role in
the discussion in this paper (because we discuss solely the
fluctuations of the $\phi$-field) as can be seen from the basic
equations expressed in the e-fold time. Eq.(16) takes the form
\begin{equation}
d\phi = -\frac{V^{\prime}}{3H^{2}}d\nu
+\frac{\gamma}{3H^{\frac{3}{2}}} a^{-\frac{3}{2}}\circ
dB+\frac{1}{2\pi}H\circ dW,\end{equation} Eq.(18)
\begin{equation}\begin{array}{l}
\partial_{\nu}P=\frac{\gamma^{2}}{18}\partial_{\phi}\frac{1}{(Ha)^{\frac{3}{2}}}\partial_{\phi}\frac{1}{(Ha)^{\frac{3}{2}}}P
+\frac{1}{8\pi^{2}}\partial_{\phi}H\partial_{\phi}HP
+\partial_{\phi}(3H)^{-2}V^{\prime}P .\end{array}\end{equation}
Eq.(45) is
\begin{equation}
P_{\infty}=(Ha)^{\frac{3}{2}}\tilde{P}
\end{equation}
where $\tilde{P}$ is defined by the same formula (46). So, the
change of time has a minor effect on the expectation values of
$\tilde{\epsilon}$ , $\tilde{\eta}$ and $\partial^{2}a$ but does
not change substantially our conclusions.


\begin{thebibliography}{99}

\bibitem{lindeinflation}
A. Linde, Phys.Lett.{\bf B129},177(1983)


\bibitem{starobinsky}A.A.Starobinsky, in Current Topics in Field Theory,Quantum Gravity and Strings, ed. By H.J.Vega and N. Sanchez, Lecture Notes in Phys.{\bf 246}, Springer, 1986


\bibitem{nelson} E. Nelson, Dynamical Theories of Brownian Motion,

Princeton Univ.Press,1967
\bibitem{bererafang}A. Berera and Li-Zhi Fang, Phys.Rev.Lett.{\bf
74},1912(1995)

\bibitem{warm}A. Berera,I.G. Moss and R.O.Ramos,

Rep.Progr.Phys.{\bf 72},026901(2009)
\bibitem{power1} M.H. Hall, I.G. Moss and A.Berera,

 Phys.Rev.{\bf D69},083525(2004)
\bibitem{power2} R.O. Ramos and L.A. da Silva, JCAP03(2013)032

\bibitem{habaejc} Z. Haba, Eur.Phys.J.{\bf
C78},596(2018) \bibitem{star1} A.D. Linde, Phys.Lett.{\bf
B116},335(1982) \bibitem{star2} A.A. Starobinsky, Phys.Lett.{\bf
B117},175(1982) \bibitem{star3} A. Vilenkin and L.H. Ford,
Phys.Rev.{\bf D26},1231(1982)
\bibitem{vilenkin}A. Vilenkin,
Phys.Rev. {\bf D27},2848(1983)

\bibitem{venn}V. Vennin, H.Assadullahi, H. Firouzjahi,
M.Noorbala and D. Wands, Phys.Rev.Lett.{\bf 118},031301(2017)
\bibitem{venn3} H.Assadullahi, H. Firouzjahi,
M.Noorbala,V. Vennin and D. Wands, JCAP06(2016)043

\bibitem{habad}Z. Haba, Int.Journ Mod.Phys.{\bf D28},1950085
(2019)
\bibitem{inflation} V. Mukhanov, Physical Foundations of
Cosmology, Cambridge,2005
\bibitem{berrera}A. Berera, Phys.Rev.{\bf
D54},2519(1996)
\bibitem{adv}Z. Haba, Adv.High.Energy.Phys.{\bf
2018},7204952(2018)

;arXiv:1807.00639

\bibitem{habaprep}Z. Haba, in preparation
\bibitem{starobquan}A.A. Starobinsky and J. Yokoyama,

Phys.Rev.{\bf D50},6357(1994)

\bibitem{ikeda} N. Ikeda and S. Watanabe, Stochastic
Differential Equations and Diffusion Processes, North Holland,1981


\bibitem{gikhman} I.I. Gikhman  and A.B. Skorohod, Stochastic Differential
  Equations,  Springer,1972





   \bibitem{ency}J. Martin, C. Ringeval and V.Vennin, Encyclopedia
   Inflationaris,    arXiv:1303.3787

\bibitem{starpot0} S.V. Ketov and A.A. Starobinsky,
JCAP08(2012)022
\bibitem{starpot}A. Kehagias, A.M. Dizgah and A. Riotto,


Phys.Rev.{\bf D89},043527(2014)

\bibitem{natural} K. Freese, J.A. Frieman and A.V. Olinto,
Phys.Rev. Lett.{\bf 65},3233(1990)

\bibitem{khasminski}R. Z. Khasminski, Stochastic Stability of
Differential

Equations, Sijthoff and Noordhoof, 1980
\bibitem{mackey}A. Lasota and M.C. Mackey, Probabilistic
Properties of Deterministic Systems, Cambridge, 1985


\bibitem{linde}A. D. Linde, Phys.Rev. {\bf
D58},083514(1998)

A. D. Linde, Phys.Rev. {\bf D49},748(1994)



\bibitem{vilenkininflation} A. Vilenkin, Nucl. Phys.,Proc.Suppl.,{\bf
B88},67(2000)
\bibitem{powerlaw}F. Lucchin and S. Mataresse, Phys.Rev.{\bf
D32},1316(1985)

\bibitem{mar}F. Finnelli, G. Marozzi, A.A. Starobinsky and G. Venturi,
Phys.Rev. {\bf D79},044007(2009)
\bibitem{starven}V. Vennin and AA. Starobinsky,

 Eur.Phys.J. {\bf C75},413(2015)




\end{thebibliography}
\end{document}